\documentclass[12pt]{article}
\usepackage{graphicx}
\usepackage{subcaption}
\usepackage{flexisym}

\newcommand\pubnumber{SNSN-323-63}
\newcommand\pubdate{\today}

\def\Title#1{\begin{center} {\Large #1 } \end{center}}
\def\Author#1{\begin{center}{ \sc #1} \end{center}}
\def\address#1{\begin{center}{ \it #1} \end{center}}

\newcommand\pubblock{\rightline{\begin{tabular}{l} \pubnumber\\
         \pubdate  \end{tabular}}}
\newenvironment{Abstract}{\begin{quotation}  }{\end{quotation}}
\newenvironment{Presented}{\begin{quotation} \begin{center} 
             PRESENTED AT\end{center}\bigskip 
      \begin{center}\begin{large}}{\end{large}\end{center} \end{quotation}}

\def\beq{\begin{equation}}
\def\eeq#1{\label{#1}\end{equation}}
\def\eeqn{\end{equation}}

\def\beqa{\begin{eqnarray}}
\def\eeqa#1{\label{#1}\end{eqnarray}}
\def\eeqan{\end{eqnarray}}

\let\bar=\overbar

\def\Dslash{\not{\hbox{\kern-4pt $D$}}}
\def\dslash{\not{\hbox{\kern-2pt $\del$}}}

\def\msb{{\bar{\ssstyle M \kern -1pt S}}}

\begin{document}
\begin{titlepage}
\pubblock

\vfill
\Title{A SOI-Based Low Noise and Wide Dynamic Range Event-Driven Detector for X-Ray Imaging}
\vfill

\Author{Sumeet Shrestha$^{1,}$, Hiroki Kamehama$^{1}$, Shoji Kawahito$^{1}$, Keita Yasutomi$^{1}$, Keiichiro Kagawa$^{1}$, Ayaki Takeda$^{2}$, Takeshi Go Tsuru$^{2}$, Yasuo Arai$^{3}$}


\address{%
$^{1}$ Research Institute of Electronics, Shizuoka University, 3-5-1 Johoku Nakaku, Hamamatsu 432-8011, Japan\\
$^{2}$ Department of Physics, Kyoto University, Kitashirakawa-Oiwake-Cho, Sakyo 606-8502, Japan\\
$^{3}$ High Energy Accelerator Research Organization (KEK), Institute of Particle
and Nuclear Studies, Oho 1-1, Tsukuba, Ibaraki 305-0801, Japan\\}
\vfill
\begin{Abstract}
A low noise and wide dynamic range event driven detector for the detection of X-Ray energy is realized using 0.2 [$\mu$m] Silicon on insulator (SOI) technology. Pixel circuits are divided into two parts; signal sensing circuit and event detection circuit. Event detection circuit is activated when X-Ray energy falls into the detector. In-pixel gain selection is implemented for the detection of a small signal and wide band of energy particle. Adaptive gain and capability of correlated double sampling (CDS) technique for the kTC noise canceling of charge detector realizes the low noise and high dynamic range event driven detector.
\end{Abstract}
\vfill
\begin{Presented}
International Workshop on SOI Pixel Detector (SOIPIX2015), Tohoku University, Sendai, Japan, 3-6, June, 2015
\end{Presented}
\vfill
\end{titlepage}

\section{Introduction}

For the scientific imaging, particularly in high energy physics, photo detectors with a fully depleted substrate and on-chip circuitry are required ~\cite{ANIMMA}\index{ANIMMA}~\cite{IISW_1} \index{IISW_1}. The SOI (Silicon-On-Insulator) pixel detector with a fully-depleted substrate as a photo detector and SOI circuits for in-pixel processing are an ideal solution for such scientific imaging for high energy physics ~\cite{ISDASTD}\index{ISDASTD}~\cite{Vertex}\index{Vertex}~\cite{IISW_2}\index{IISW_2}. Also detector with wide bandpass and superior hit position readout of 10 [$\mu$s] is required. This paper proposes a novel SOI pixel detector with potential profile for charge collection back-gate surface potential pinning for improving the sensitivity, stable operation of SOI circuits and low noise. Pixel circuit and readout circuit can create bottleneck problem to realize the full advantages of proposed detector. This papers describes the event-driven pixel circuits for improving low noise performance and wide dynamic range. 

\section{Detector Structure}

\begin{figure}[htb]
\centering
\includegraphics[height=2in]{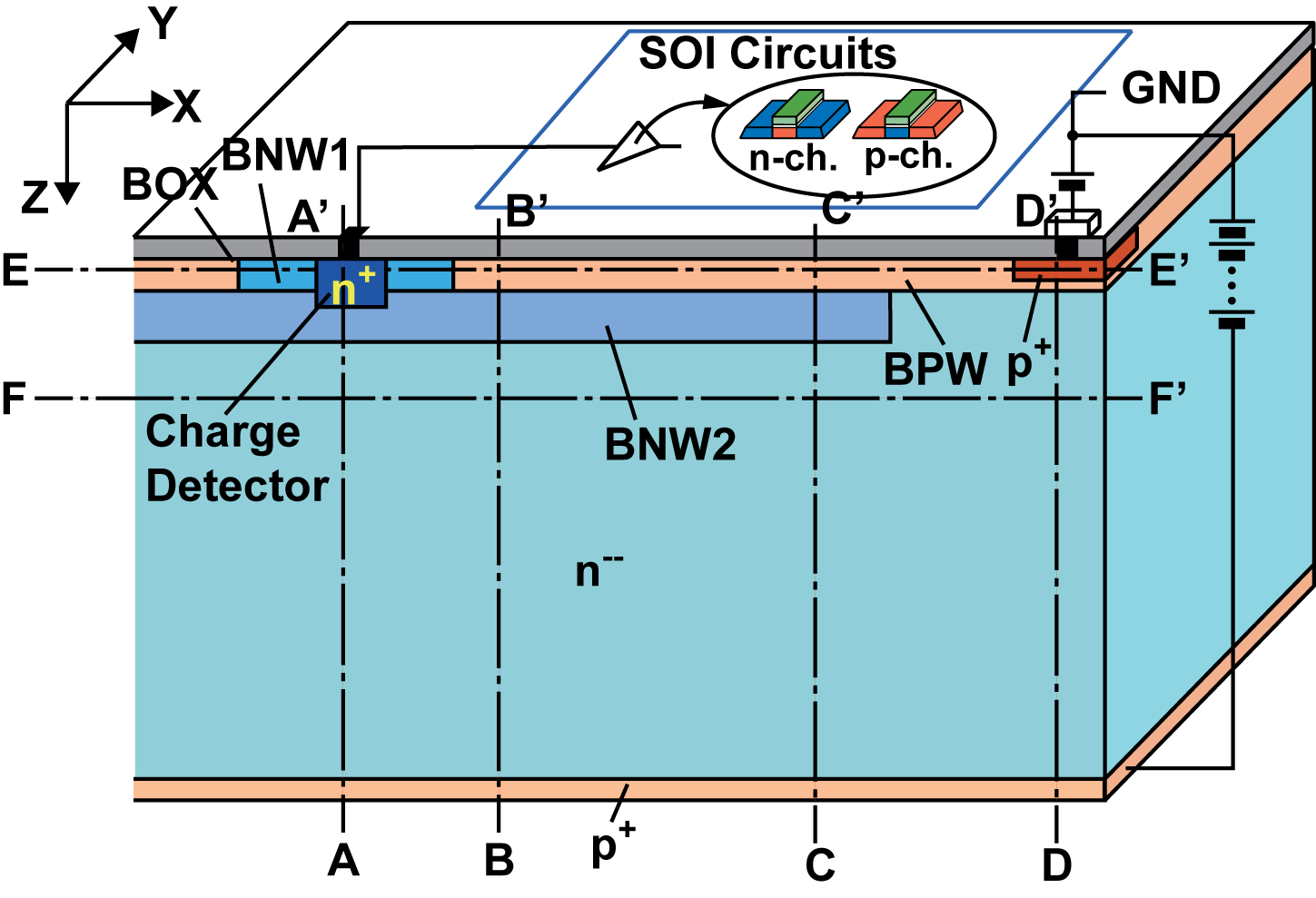}
\caption{Proposed Detector.}
\label{fig:Detector}
\end{figure}

Fig.\ref{fig:Detector} shows a conceptual schematic of the proposed SOI pixel detector. In order to reduce the back-gate effect to SOI circuits, a BPW (Buried P-Well) is formed underneath the SOI circuits. A charge collector n+ and two different buried n-wells, BNW1 and BNW2 are formed in the n-type high-resistivity substrate. p+ layer is formed at the backside of the substrate with a back-end process steps. The BNW2 plays an important role for creating lateral electric field for high-speed charge collection and increasing a potential barrier to holes in the BPW. This allows us to use this detector under a fully depleted condition by applying negative voltage to the backside p+ region while preventing the punch-through to the back-gate (BPW) and injection of holes from the BPW. Since the BNW1 and BNW2 are fully depleted and the capacitance of the charge collector is only due to the n+/BPW junction, high charge-to-voltage conversion gain is realized.

\section{Event-driven Pixel Circuit}

\begin{figure}[htb]
\centering
\includegraphics[height=4in]{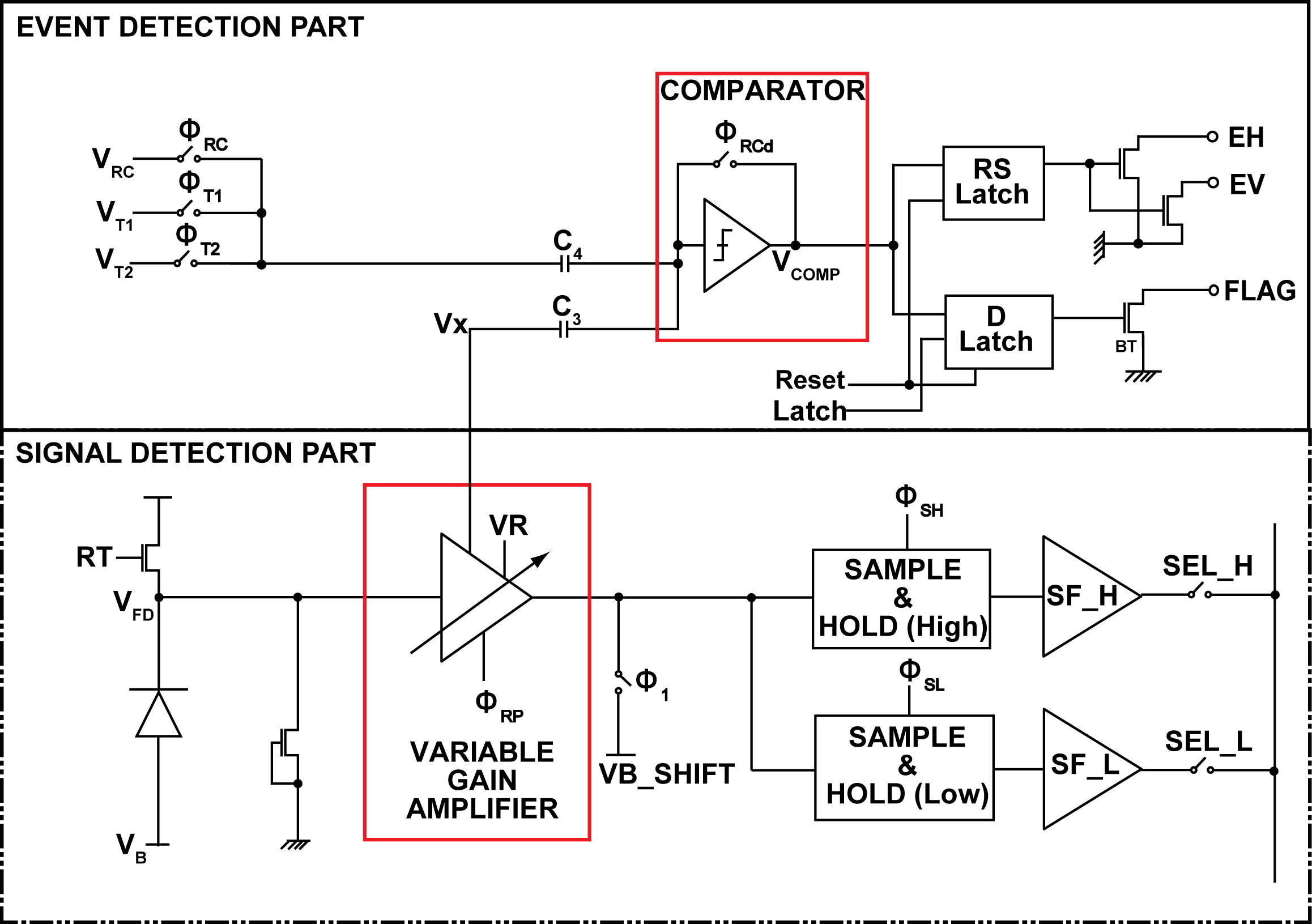}
\caption{Event-driven Pixel Circuit.}
\label{fig:pixel_ckt}
\end{figure}

Fig.\ref{fig:pixel_ckt} shows the event-driven pixel circuit. Pixel circuit is functionally divided into two parts: Signal sensing circuit and event detection circuit. During the reset phase, reset signal (Vin(R)) is sampled. \\

Event signal (EH, EV) from a event detector circuit is continuously scanned. When the detector absorbs the X-ray energy (Vin(S)), an event is triggered given by equation 1 where V$_{T1}$ is the minimum threshold voltage for an event to occur.
\begin{equation}
Event = 
\left\lbrace
\begin{tabular}{ll}
``1" &if $\left(V_{in}(S) - V_{in}(R) + \frac{C_3}{C_4}V_{T1}\right) \leq 0$\\
``0" &if $\left(V_{in}(S) - V_{in}(R) + \frac{C_3}{C_4}V_{T1}\right) >0$
\end{tabular}
\right.
\end{equation}

After the event is detected, threshold voltage is changed from V$_{T1}$ to V$_{T2}$ for the evaluation of the signal strength. Flag signal is used for the indication of the signal strength as given by equation 2. Gain is then selected depending upon the flag signal. 
\begin{equation}
Flag = 
\left\lbrace
\begin{tabular}{ll}
``1" &if $\left(V_{in}(S) - V_{in}(R) + \frac{C_3}{C_4}V_{T2}\right) \leq 0$\\
``0" &if $\left(V_{in}(S) - V_{in}(R) + \frac{C_3}{C_4}V_{T2}\right) >0$
\end{tabular}
\right.
\end{equation}

\begin{figure}[htb]
\centering
\includegraphics[height=2in]{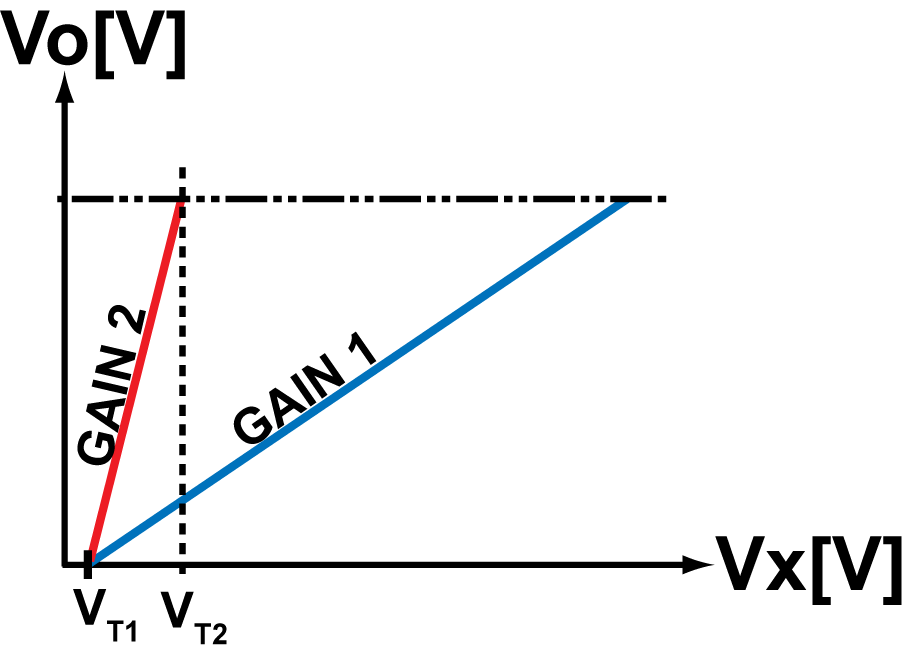}
\caption{Output response for different gain.}
\label{fig:gain_vt1_vt2}
\end{figure}

Fig.\ref{fig:gain_vt1_vt2} shows the expected output response for two different gain where GAIN_2 is greater then GAIN_1. High gain is provided for a low energy spectrum and small gain of 1 or 2 is provided for a high energy spectrum. 

\section{Simulation Results}
\begin{figure}[htb]
\begin{subfigure}{.5\textwidth}
  \centering
  \includegraphics[width=.9\linewidth, height=1.8in]{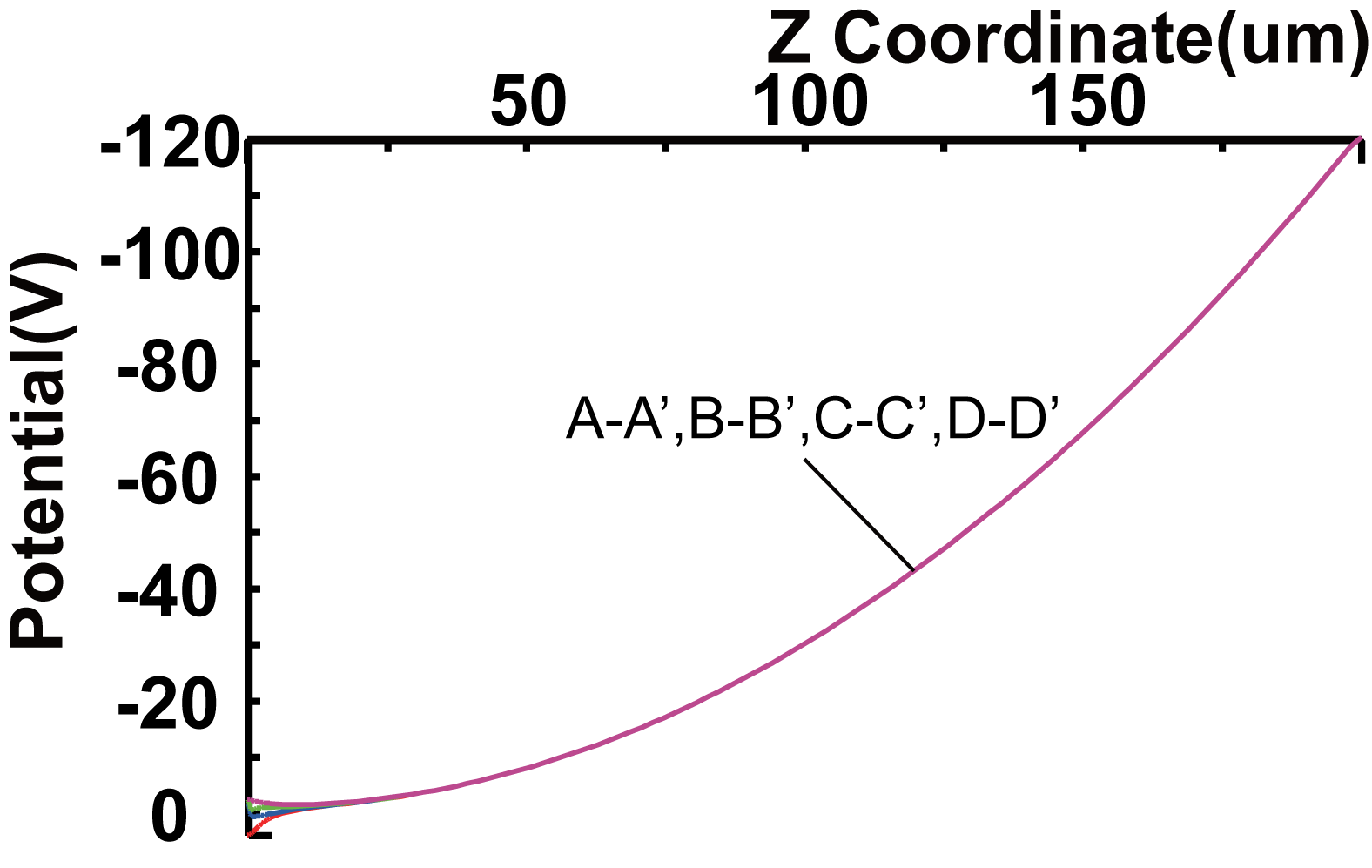}
  \caption{2a}
  \label{fig:TYP7_ABCD}
\end{subfigure}
\begin{subfigure}{.5\textwidth}
  \centering
  \includegraphics[width=.9\linewidth, height=1.8in]{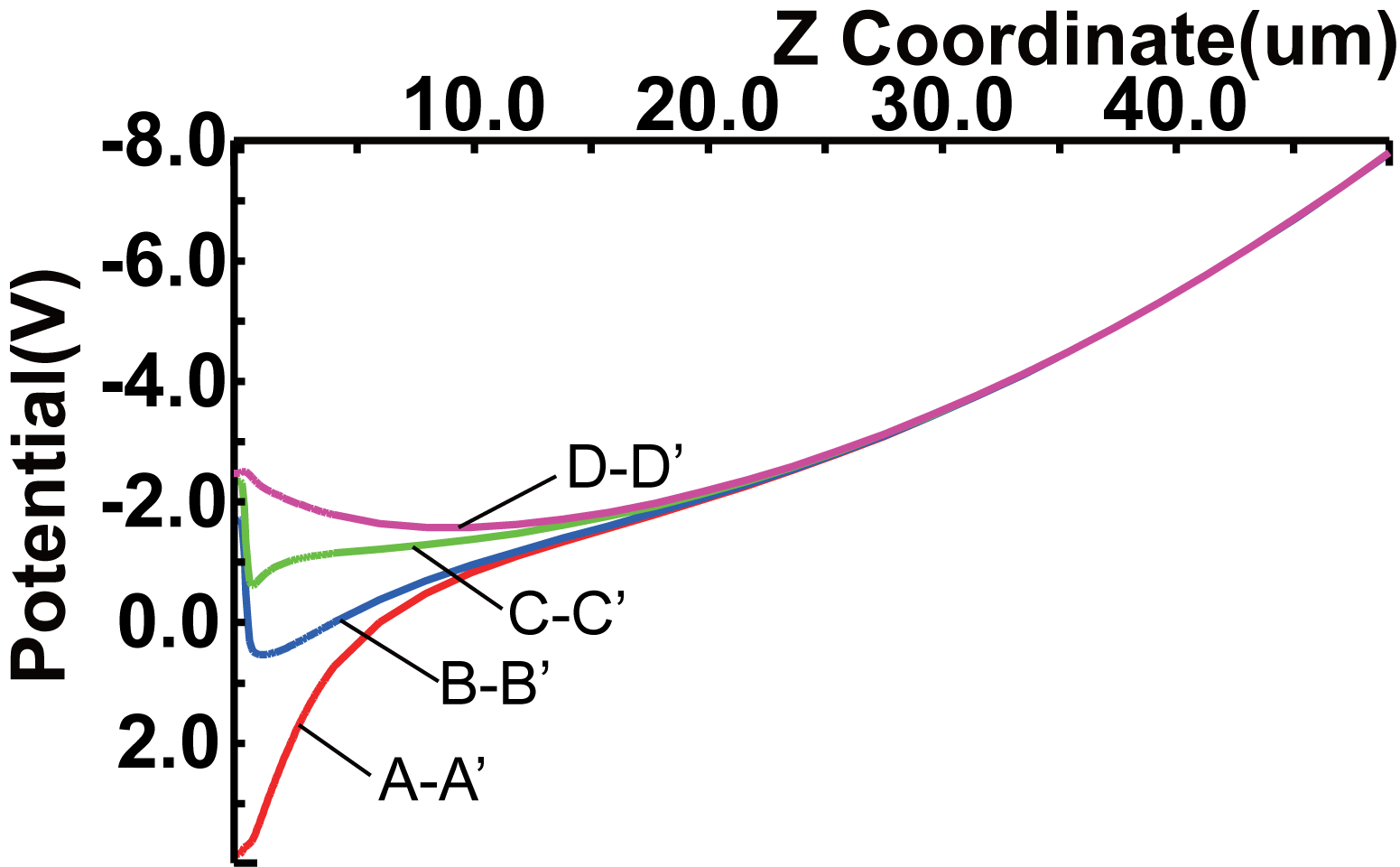}
  \caption{2b}
  \label{fig:TYP7_VSUB-120V_ABCD}
\end{subfigure}
\caption{Potential distribution along vertical cross-section.}
\label{fig:V_Pot_Dis}
\end{figure}

\begin{figure}[htb]
\centering
\includegraphics[height=1.5in]{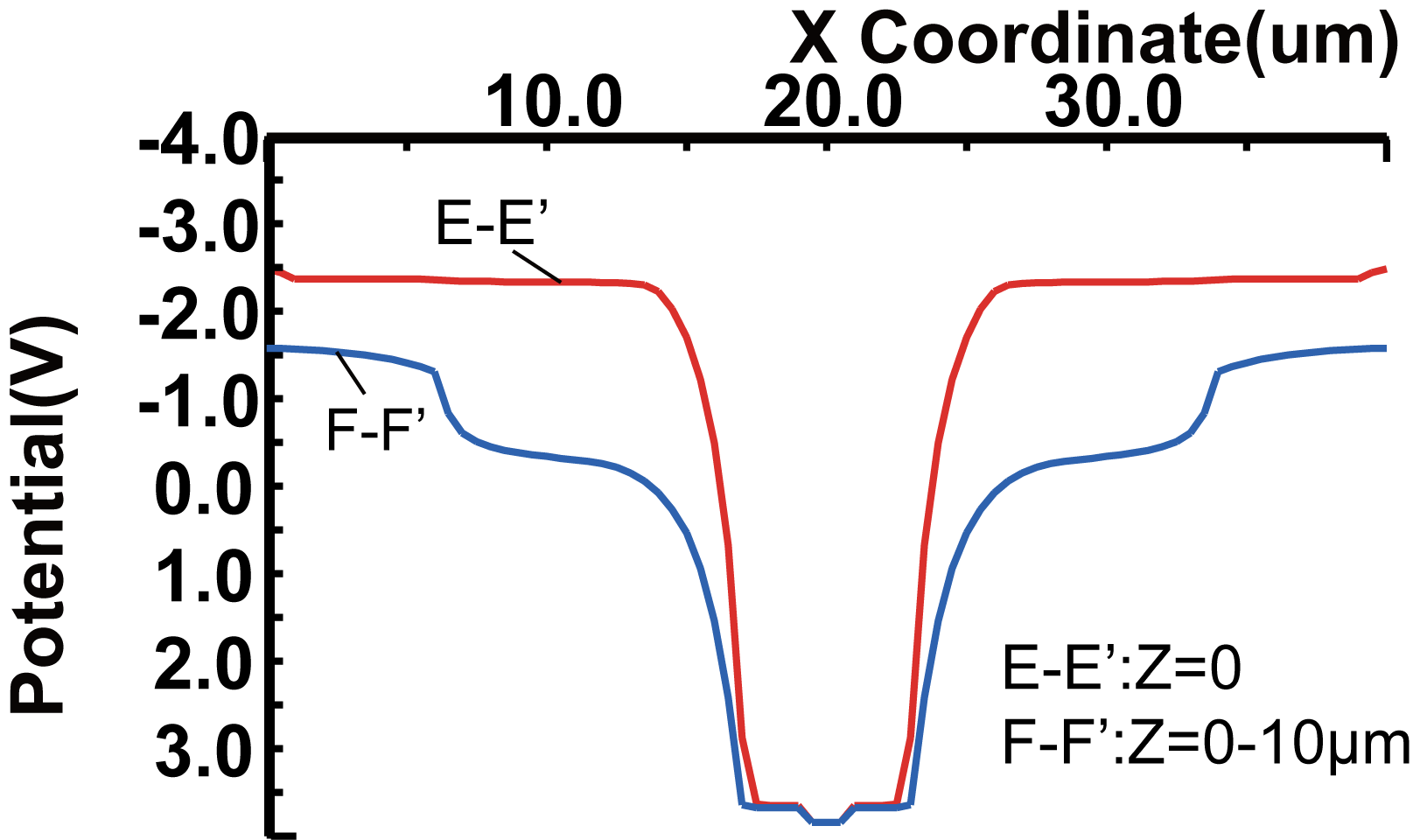}
\caption{Potential distribution along horizontal cross-section.}
\label{fig:H_Pot_Dis}
\end{figure}

Fig.\ref{fig:V_Pot_Dis} and Fig.\ref{fig:H_Pot_Dis} shows the simulated potentials of the SOI pixel detector. The pixel size of the detector is 40$\mu$mX40$\mu$m,and thickness of sensor-layer (n$-$substrate) is 200 [$\mu$m]. The voltages set at the charge detector, the back-gate(BPW), and the substrate backside p+ (Vback) are 3 [V], -2 [V] and -120 [V], respectively. Fig.\ref{fig:TYP7_ABCD} shows potential distribution of vertical cross-sections of A-A\textprime, B-B\textprime, C-C\textprime and D-D\textprime. The entire sensor layer is fully depleted from the backside to the surface. Fig.\ref{fig:TYP7_VSUB-120V_ABCD} is a zoomed potential distribution from the depth of 50 [$\mu$m] to the surface.\\

In the cross-sections of B-B\textprime, C-C\textprime and D-D\textprime, the back-gate surface is pinned to the supply voltage of the BPW(-2 [V]), while creating a potential distribution and resulting vertical and lateral electric fields to accelerate photo-electrons to the n$+$ charge collector.BNW2 creates a large potential barrier to holes in the BPW and is almost unchanged for the variation of the backside junction voltage. The punch-through is prevented even if an excess bias voltage (-130 [V] to -140 [V]) is applied. As shown in Fig.\ref{fig:H_Pot_Dis} which is a horizontal maximum potential distribution at the cross-section of E-E\textprime and F-F\textprime, the back-gate (BPW) is pinned to -2.4 [V] and lateral electric field to collect photo-electrons in the pixel to the charge collector is created. \\

\begin{figure}[htb]
\centering
\includegraphics[height=2.2in]{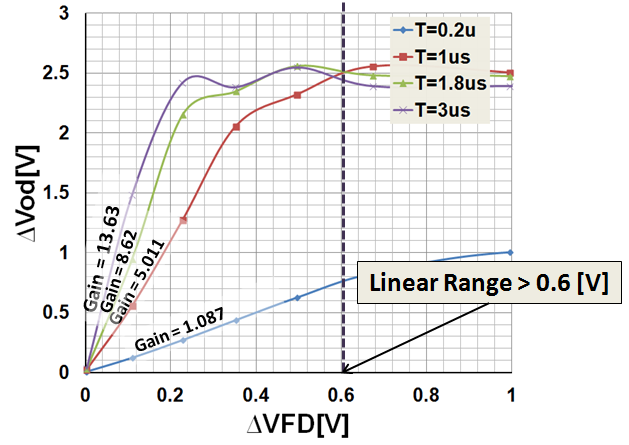}
\caption{Output characteristics of pixel circuit.}
\label{fig:pixel_sim}
\end{figure}

Fig.\ref{fig:pixel_sim} shows the simulation result for the event-driven pixel circuit. VFD is the input voltage at floating diffusion node and Vod is the output at VH or VL node. Time is varied from 0.2 [$\mu$s] to 3 [$\mu$s] and the response of the pixel output was observed. Gain can be changed from 1 to 13. If the incident energy spectrum is high enough we provide the gain of 1 and if the incident energy has very low energy spectrum high gain (e.g. 13) is provided.  

\begin{figure}[htb]
\centering
\includegraphics[height=2.2in]{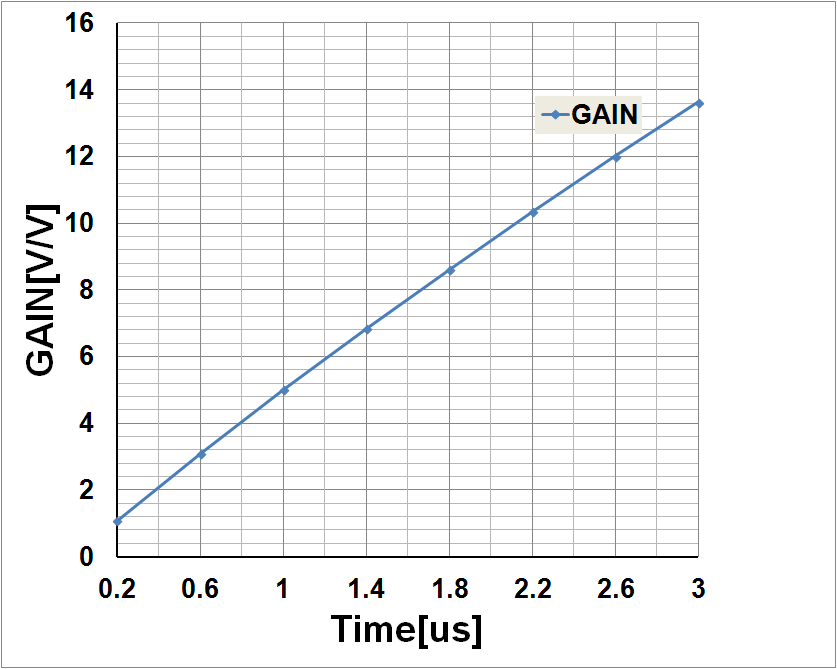}
\caption{Gain linearity curve.}
\label{fig:Time_Vs_Gain}
\end{figure}

Fig.\ref{fig:Time_Vs_Gain} shows the gain linearity curve. Gain linearly increase with the increase in time. In-pixel gain selection is used to select two different gain for low and high energy spectrum.

Thus, detector with high stability, large conversion gain and high charge collection efficiency was realized. Also, pixel circuit for low noise and high dynamic range was developed.

\end{document}